\documentclass[conference]{IEEEtran}
\IEEEoverridecommandlockouts
\usepackage{graphicx}
\usepackage{epsf}
\usepackage{epstopdf}
\usepackage{cite}
\usepackage{amsmath}
\usepackage{amssymb}
\usepackage{color}

\ifCLASSINFOpdf

\else

\fi

\begin{document}

\title{Kinetic model of selectivity and conductivity of the KcsA filter }

\author{
\IEEEauthorblockN{W.A.T. Gibby}
\IEEEauthorblockA{Department of Physics\\ Lancaster University\\
	 Lancaster LA1 4YB, UK \\
	 Email: w.gibby@lancaster.ac.uk}
	 \and 
	 \IEEEauthorblockN{D.G. Luchinsky} 
\IEEEauthorblockA{SGT Inc., Greenbelt\\ MD, 20770, USA\\
}\and 

 \IEEEauthorblockN{I.Kh. Kaufman\\A. Ward\\ and  P.V.E.
	McClintock}
\IEEEauthorblockA{Department of Physics\\ Lancaster University\\
	 Lancaster LA1 4YB, UK }}

\maketitle

\begin{abstract}
We introduce a self-consistent multi-species kinetic theory based on the structure of the narrow voltage-gated potassium channel. Transition rates depend on a complete energy spectrum with contributions including the dehydration amongst species, interaction with the dipolar charge of the filter and, bulk solution properties.  It displays high selectivity between species coexisting with fast conductivity, and Coulomb blockade phenomena, and it fits well to data. 
\end{abstract}

\IEEEpeerreviewmaketitle
%\documentclass[conference]{IEEEtran}
%\IEEEoverridecommandlockouts
%\usepackage{graphicx}
%\usepackage{cite}
%\usepackage{epsf}
%\usepackage{epstopdf}
%\usepackage{amsmath}
%\usepackage{amssymb}
%\usepackage{color}
%\ifCLASSINFOpdf
%
%\else
%
%\fi
%\hyphenation{op-tical net-works semi-conduc-tor}
%
%
%\begin{document}
%
%\title{\textbf{}}
%

%\IEEEpubid{\makebox[\columnwidth]{\hfill 978-1-5090-2760-6/17/\$31.00 \copyright~2017 IEEE}
%	\hspace{\columnsep}\makebox[\columnwidth]{Published by the IEEE Computer Society}}
%
%
%
%
%

%
%\maketitle
%
%
%% 
%
%\IEEEpubid{\makebox[\columnwidth]{\hfill 978-1-5090-2760-6/17/\$31.00 \copyright~2017 IEEE}
%	\hspace{\columnsep}\makebox[\columnwidth]{Published by the IEEE Computer Society}}
%
%\IEEEpeerreviewmaketitle

\section{Introduction}

Narrow biological ion channels allow the permeation of ions at near to the diffusion rate, whilst being highly selective.  Paradoxically, potassium (K$^+$) channels allow conduction at close to 10$^8$s$^{-1}$, coexisting  with high selectivity amongst mono-valent species \cite{Hille:01}.  

The general permeation properties through the filter are preserved among all voltage-gated K$^+$ channels due to the conserved amino acid structure of the filter \cite{Kuang:15}.  Experimentally it is known that conduction occurs via the single-file concerted motions of ions separated by a water molecule, such that we have transitions between 2 and 3 ions \cite{Hodgkin:55,MacKinnon:2001b}. The filter charge $Q_f = n_fq$ plays an important role in the permeation process. It is attributed to the dipolar interactions with the oxygen atoms in the permeation pathway, and these 20 atoms form 5 binding sites within the filter S1-S4 and 1 site in the entrance to the extracellular bulk S0.

%Structure of potassium channels
%Qie Kuang1,2 • Pasi Purhonen1 • Hans Hebert1,2

Kinetic theory has  been widely studied and applied to channels    \cite{Lauger:73,Kitzing:92,Tolokh:06,Nelson:11}, with some success in comparing  predictions with experimental data of K$^+$ channels  \cite{Heginbotham:93,Meuser:99} and explaining properties such as rectification. Yet kinetic models are often criticised for inconsistencies in the definition of the transition rates and the choice of energy barriers \cite{Cooper:88}.

Here we introduce a novel self-consistent kinetic model of the KcsA filter that resolves its selectivity $vs$ conductivity paradox. The model is based on a first principles statistical analysis of the energy of the filter that incorporates the effects of site-binding, dehydration, excluded volume etc by including the excess chemical potential difference between the bulk and the channel $\Delta \bar{\mu}^b_i=\bar{\mu}^b_i-\bar{\mu}_i^c$ \cite{Luchinksy:16}. The self-consistent transition rates are included into the model using Grand Canonical Monte Carlo theory \cite{Roux:1999}. We validate the theory by comparing model predictions with experimental data, including mixed species solutions and mutants \cite{Heginbotham:93,zhou:2004}.

\begin{figure}
\begin{center}
\includegraphics[width = 8cm, height = 5.5cm]{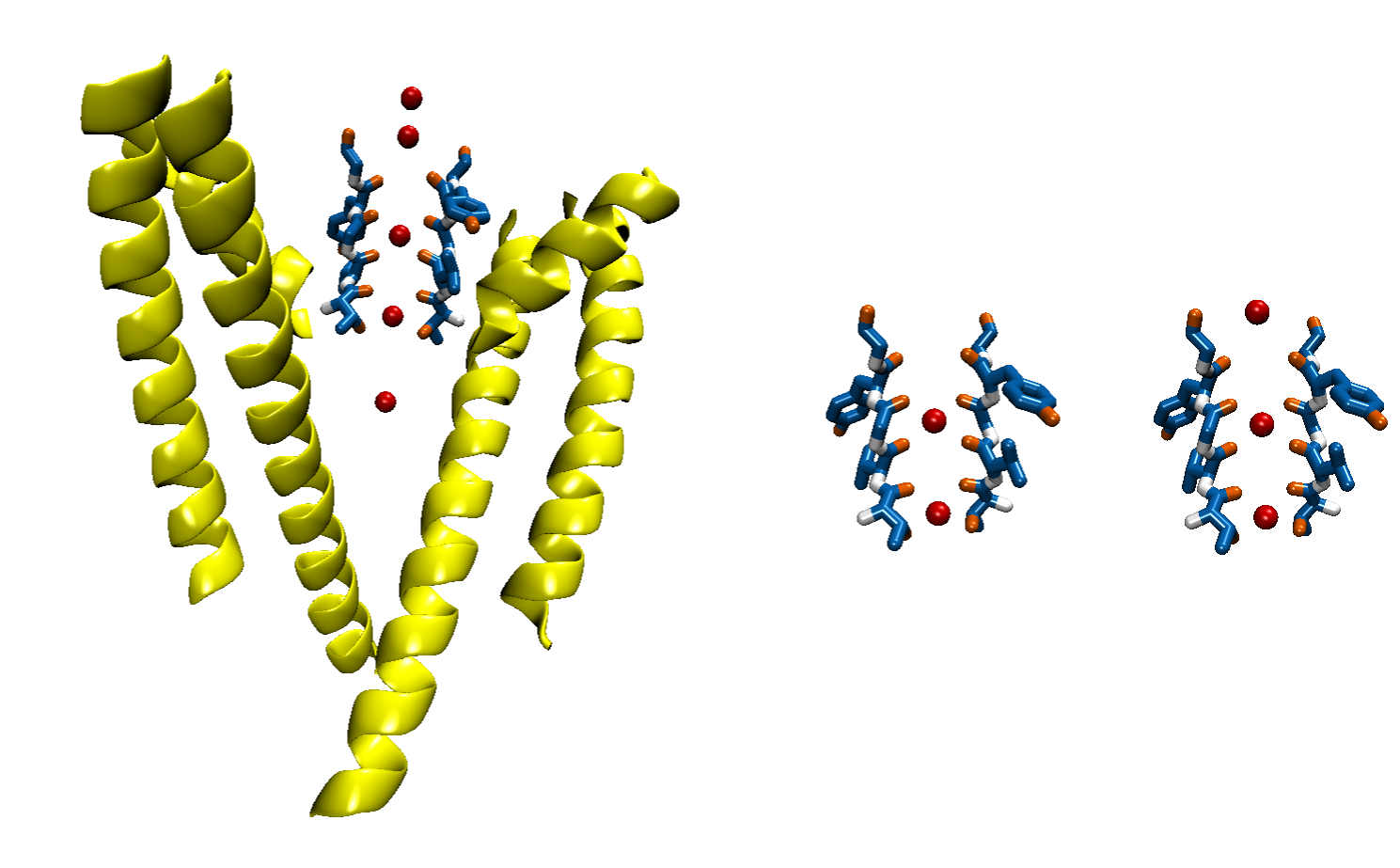}
\caption{\textbf{Left:} reduced image of closed conformation high-K$^+$ KcsA (pdb 1K4C), rendered with VMD software \cite{HUMP:96}. The yellow ribbons represent two of the amino acid chains, whilst the narrow selectivity filter has its residues highlighted. The 4/5 binding sites are displayed with a S0S2S4 occupation. \textbf{Right:} Schematic of the transition S2S4$\leftrightarrow$S0S2S4 (note that we may also have an additional two ion state S1S3).   }\label{Fig:Model}
\end{center}
\end{figure}

In the work that follows, with SI units: $q, k,$ $N_0$, $T$, $\epsilon_0$ and  $\epsilon_w$, respectively represent the proton charge, Boltzmann's and Avogadro's constants, system temperature, and water and vacuum dielectric permittivities ($\epsilon_w=80$) and  constants. We use the following bulk diffusion coefficients:  $D_K^b=1.96\times 10^{-9}$ m$^2$s$^{-1} $  and $D_{Na}^b=1.33\times 10^{-9}$ m$^2$s$^{-1} $, and ionic radii: $R_K=1.4$ \AA \; and $R_{Na}=1$\AA.  The filter has the following geometry, radius 1.5 \AA, length 12\AA, and  5 binding sites.

\section{Mixed-species Kinetic theory}

We consider a filter diffusively and thermally coupled to intra-cellular ($L$), and extra-cellular ($R$) bulk reservoirs, between which the membrane potential ($\phi_m$) acts from left to right. Anions cannot enter the filter and so we only account for their presence via their contributions to the energy spectrum.

We introduce  indistinguishable sites, with the following reduced state space describing the number and species of occupying ions $\{n_j\} $,

\begin{equation}
\{K^+K^+\}, \{K^+K^+K^+\}, \{Na^+K^+K^+\}.
\end{equation}To simplify notation the subscript 0 will denote the ground state, whilst either excited state will be distinguished by its species:  K$^+$ or  Na$^+$, denoted by $i$.  Transitions are described via a set neighbouring states master-equations, 

\begin{equation}\begin{pmatrix} 
\dot{P}_{0} \\
\dot{P}_{K} \\
\dot{P}_{Na} \\
\end{pmatrix}=
\begin{pmatrix}
-\Gamma_{01}^K-\Gamma_{01}^{Na} & 
\Gamma_{10}^K&
\Gamma_{10}^{Na} \\
\Gamma_{01}^K & 
-\Gamma_{10}^K&
0\\
\Gamma_{01}^{Na} &
0& 
-\Gamma_{10}^{Na}
\end{pmatrix} \cdot \begin{pmatrix}
\dot{P}_{0} \\
\dot{P}_{K} \\
\dot{P}_{Na} \\
\end{pmatrix}, \label{Eqn:Multi-ionMaster}
\end{equation}where we have introduced the notation: $\Gamma^i=\Gamma^{L,i}+\Gamma^{R,i}$. The general solution can be found using standard linear algebra methods when we consider probability conservation, taking a simplified form by considering the binding factor $\mathcal{B}$ \cite{Roux:1999} defined via the ratio of probabilities to the ground state,

\begin{equation}
P(\{n_j\})=\frac{\mathcal{B}(\{n_j\})}{\sum_{\{n_j\}}\mathcal{B}(\{n_j\})}, \quad \mathcal{B}(\{n_j\})=\frac{P(\{n_j\})}{P_0}.
\end{equation} To proceed in analysing the theory we first need to derive our rates, and we follow \cite{Im:00,Roux:1999,Roux:04}, and use our statistical theory. This approach starts by considering transitions between the mouth, which is in quasi-equilibrium with the bulk solution and the channel. We establish the local  balance relation between each bulk $b$ and the channel, and normalise such that rates sum to 1, leaving  sigmoidal rates as functions of the full transition energy barrier: $\Delta G(\{n_j\};n_f;b) $.  Each rate is a function of the energy barrier from each bulk and so we allow the condition of  equilibrium between both bulks to be broken. This ensures we have a non-equilibrium steady-state with non-zero current and  failure to  establish the Boltzmann ratio.

\begin{align}
&\Gamma_{10}^{b,i}=\tau_i \times D^b_i/L^2\frac{\exp[(\Delta G(\{n_j\}; n_f; b))/kT]}{1+\exp[(\Delta G(\{n_j\}; n_f; b))/kT]}, 
\nonumber \\ & 
\Gamma_{01}^{b,i}=\tau_i\times D^b_i/L^2 \frac{1}{1+\exp[(\Delta G(\{n_j\}; n_f; b))/kT]}
\end{align}The prefactor $\tau_i\times D_i^b/L^2$, is added under the condition that $0<\tau_i \leq 1$. These rates converge to a Kramer's exponential with a large energy barrier, and $1/2 \tau_i\times D_i^b/L^2$ at the barrier-less  condition ($\Delta G\approx 0$) because the net flow is zero \cite{Gardiner:02}.  

The free energy barrier was derived and defined earlier in \cite{Luchinksy:16}, and here we present only a brief discussion. 

\begin{align}
\Delta G(\{n_j\}; n_f; b)&= \Delta \mathcal{E}(\{n_j\}; Q_f)  - \Delta\phi^b - kT\ln(x_i^b)
\nonumber \\ &
+kT\ln[(n_i+1)/n_w] -\Delta \bar{\mu}^b_i. \label{Eqn:EnergyBarrier}
\end{align}It represents the energy required for a transition between the bulk and channel to occur. It consists of an electrostatic component $\Delta \mathcal{E}(\{n_j\}; Q_f)$ between ions and $Q_f$ \cite{Luchinksy:16,Kaufman:15}, bulk parameters: mole fraction $\sim c_i^b/c_w^b$ (where $c$ is the concentration of either species or water molecule) and influence from the membrane potential $\Delta\phi^b$ in transition to a site at $\eta$. We  include a permutation factor due to binding statistics where $n_i$ and $n_w$ are the initial numbers of ions and water molecules in the filter, and the excess chemical potential difference between bulk and channel $\Delta \bar{\mu}^b_i$.

This final term provides a distinction between species because it depends strongly on the ionic concentration, charge and radii.  These contribute via a large number of terms including: dehydration at either filter entrance, site bonding, ion-ion interactions in the bulk and volume exclusion \cite{Luchinksy:16,McQuarie:76,Krauss:11}.  Our term takes the following form,

\begin{equation}
\Delta \bar{\mu}_i^b= \Delta \tilde{\mu}_i^b-\frac{q^2\kappa}{8\pi\epsilon_w\epsilon_0(1+\kappa R_i)},  \kappa=\sqrt{\frac{N_0 q^2\sum_i 2z_i^2c_i^b}{kT\epsilon_0\epsilon_w}}
\end{equation}where the term: $\Delta \tilde{\mu}_i^b$ includes dehydration, site-bonding and volume exclusion, and will be used as a fitting parameter. The final term is calculated explicitly  and given by the Debye-H{\"u}ckel ion-ion interaction, screened by a solvent and includes the presence of anions.  The difference of $\Delta \bar{\mu}_i^b$ between two species  gives the thermodynamic selectivity $\Delta\Delta \mu^b$, and for K$^+$ and Na$^+$ this is around $\sim 8kT$ \cite{Noskov2007a,Lockless:15}, strongly disfavouring Na$^+$.

In the next section we introduce equations for K$^+$ and Na$^+$ current and consider selectivity.

\section{Selectivity $vs$ conductivity of the KcsA filter} 

\begin{figure}[h]
\begin{center}
\includegraphics[scale=0.585]{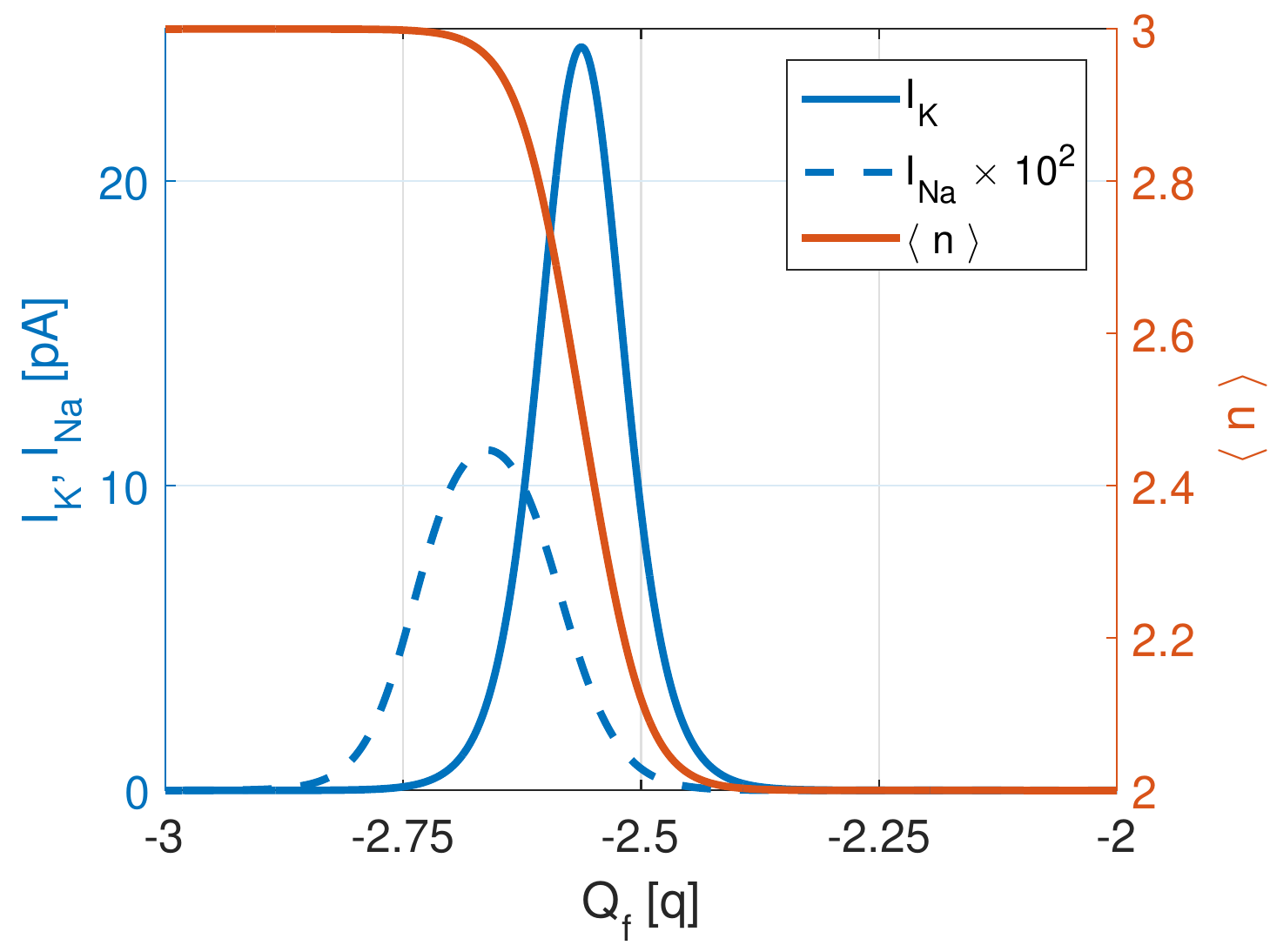}
\includegraphics[scale=0.585]{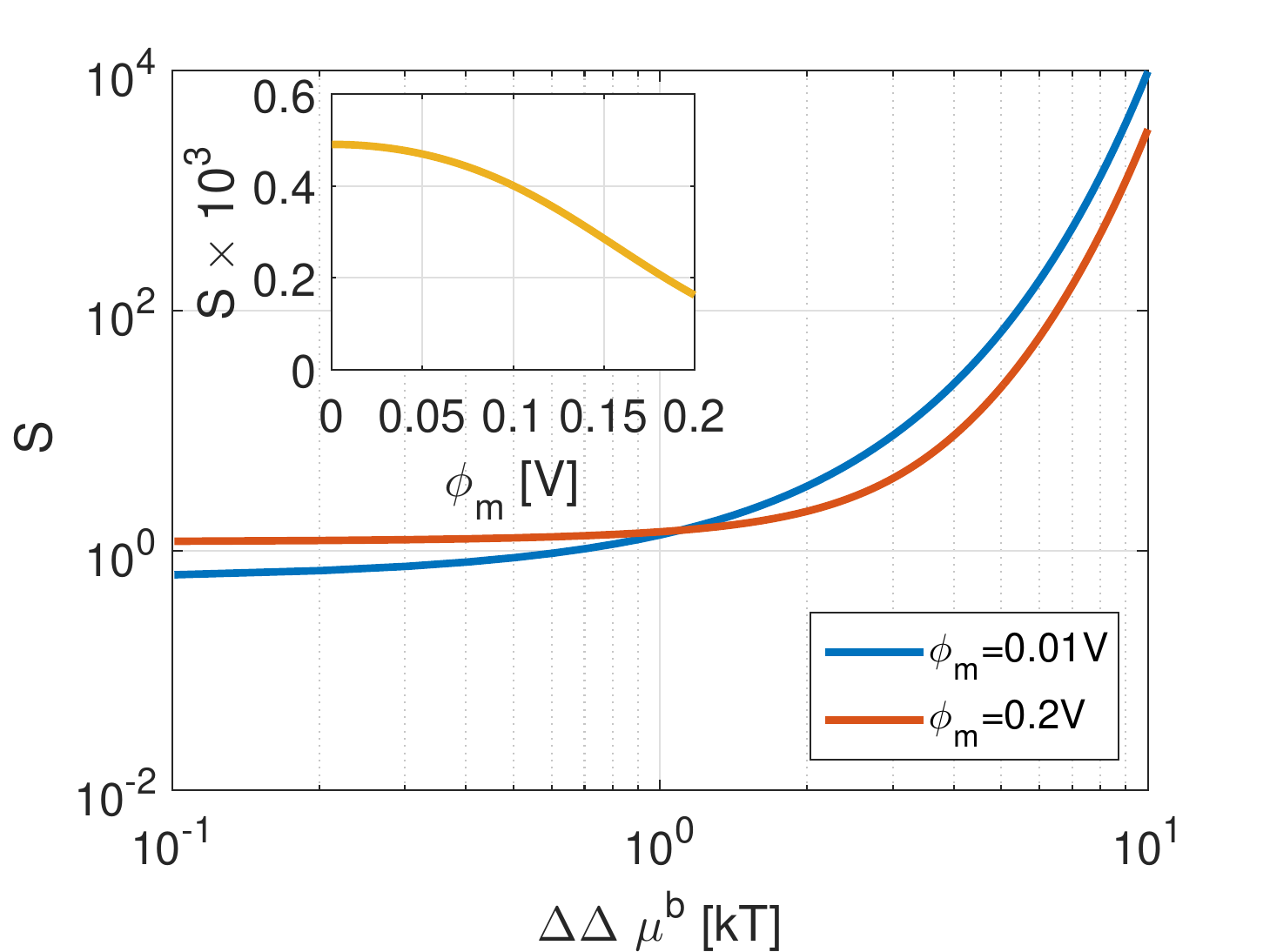}
\caption{ \textbf{Top: } $I_K$ and $I_{Na} \times 10^2$ $vs$ $Q_f$, with: $\tau_i=0.5$, $\phi_m=50$mV, $\eta=0.5$, $\Delta \tilde{\mu}_K^b=4kT$ and $x_i=0.1/55$. Each species provides a peak although Na$^+$ is shifted and $<<I_K$. \textbf{Bottom} Selectivity between species decreases $vs$  $\Delta\Delta\mu^b$ with parameters: $\tau_i=0.5$,  $Q_f$=-2.5$q$, $\eta=0.5$, $\Delta \tilde{\mu}_K^b=4kT$ and $x_i=0.1/55$. For $\Delta\Delta\mu^b \lesssim 1kT$ the filter disfavours K$^+$ and this is due to the binding statistics, for values $\gtrsim 1.1kT$ we see high selectivity. The inset plot shows selectivity decreasing $vs$ $\phi_m$ at $\Delta\Delta\mu^b=8kT$, culminating with $S\sim 160$ at $\phi_m=0.2$V.   }\label{Fig:Current}
\end{center}
\end{figure}

The current can be computed in the standard way as the balance of probability fluxes over each barrier, subject to Kirchoff's law where $I_i=I_i^L=I^R_i$. It is given below for the left barrier,

\begin{equation}
I_i=q \cdot \left(\Gamma_{01}^{L,i}\cdot P_0-\Gamma_{10}^{L,i}\cdot P_i\right).
\end{equation}After inserting the solutions to the master equations \eqref{Eqn:Multi-ionMaster}, it can be simplified to,

\begin{align}
&I_K=
    q \cdot \frac{D^c_K}{L^2} \cdot \left( \Gamma_{01}^{L,K}  -\Gamma_{01}^{R,K} \right)\cdot\frac{1}{2\frac{D^c_K}{L^2} +\Gamma_{10}^{K}\frac{\Gamma_{01}^{Na}}{\Gamma_{10}^{Na}}} \\ &
 I_{Na}  =q \cdot \frac{D^c_{Na}}{L^2} \cdot \left( \Gamma_{01}^{L,Na} - \Gamma_{01}^{R,Na} \right)\cdot\frac{1}{2\frac{D^c_{Na}}{L^2}+\Gamma_{10}^{Na}\frac{\Gamma_{01}^K}{\Gamma_{10}^{K}}}.
\end{align}This system is easily reducible back to two states describing K$^+$,  if we take the Na$^+$ concentration to be zero.  The occupancy of the filter can be calculated from the   ensemble average,

 \begin{equation}
 \langle n \rangle =2\times P_0+3\times (P_K+P_{Na})
 \end{equation}In Fig: \ref{Fig:Current} we display current and occupancy of the filter for both species  $vs$ $Q_f$. K$^+$ is energetically favoured and so its conductivity and contribution to the occupancy is much larger. Hence the  resonant transition occurs at the midpoint of the step which corresponds to a degeneracy in the energy spectrum i.e. the barrier-less knock-on condition $ \Delta G_K=0$, because here the ground and K$^+$ excited state probabilities are non-zero and equal to each other. The Na$^+$ peak is  shifted because it is trying to enter a stable fully-occupied filter and hence requires a K$^+$ ion to exit; hence it peaks at the minimum in energy to remove a K$^+$ and add a Na$^+$.   This phenomena is known as Coulomb blockade (CB) \cite{Beenakker:91,Kaufman:15}  because outside of the $K^+$ peak the filter is in a blockaded state with minimal conduction.

 Selectivity is defined  via the ratio: $S=I_K/I_{Na}$, and is plotted in the bottom figure of Fig: \ref{Fig:Current} and its inset. As expected selectivity strongly depends on the difference $\Delta\Delta\mu^b$, but decreases with membrane potential. We also observe that, due to the binding factor when $\Delta\Delta\mu^b \lesssim 1.1kT$, it results in selectivity favouring Na$^+$.  The membrane voltage provides the driving force governing conduction in symmetrical solutions, and we observe later that at large voltages the current saturates. Thus when $\phi_m\sim 0.2$V  K$^+$ conduction has, or is approaching, saturation at its diffusion limit, and it becomes independent of voltage; meanwhile  the energy barrier for Na$^+$ conduction is lowered allowing conduction. Thus at these large voltages we find a non-negligible Na$^+$ current which continues to increase with even larger $\phi_m$ until it reaches its diffusion limit. This non-negligible current is observed experimentally  and known as \textit{punch-through}  \cite{Nimigean:02}.

\section{Experimental comparisons} 

There is a vast experiential literature about the behaviour of K$^+$ channels and in this paper we will compare the theoretical current with  current-voltage ($I-V$) curves under differing concentrations, mixed solutions and the effect of filter mutations, from the Shaker channel \cite{Heginbotham:93}, KcsA \cite{zhou:2004}. Unless stated, only  K$^+$ is  present in the solution, and hence the state space reduces to: $\{K^+K^+\}, \{K^+K^+K^+\}$, and we take $Q_f=-2.5q$.

%If we consider the current at linear response then the ratio becomes,
%
%\begin{equation}
%\mathcal{S}= \frac{D^c_K/D^c_{Na}\exp[(\Delta \Delta \mu^e)/kT]}{3+\exp[\Delta\bar{\mu}_K^e-\Delta \mathcal{E}/kT]}
%\end{equation}
%
%where we have dropped the superscripts denoting bulks because we are at equilibrium, and the factor of 3 arises from our permutations term. 

\begin{figure}[h!]
\begin{center}
\includegraphics[scale=0.585]{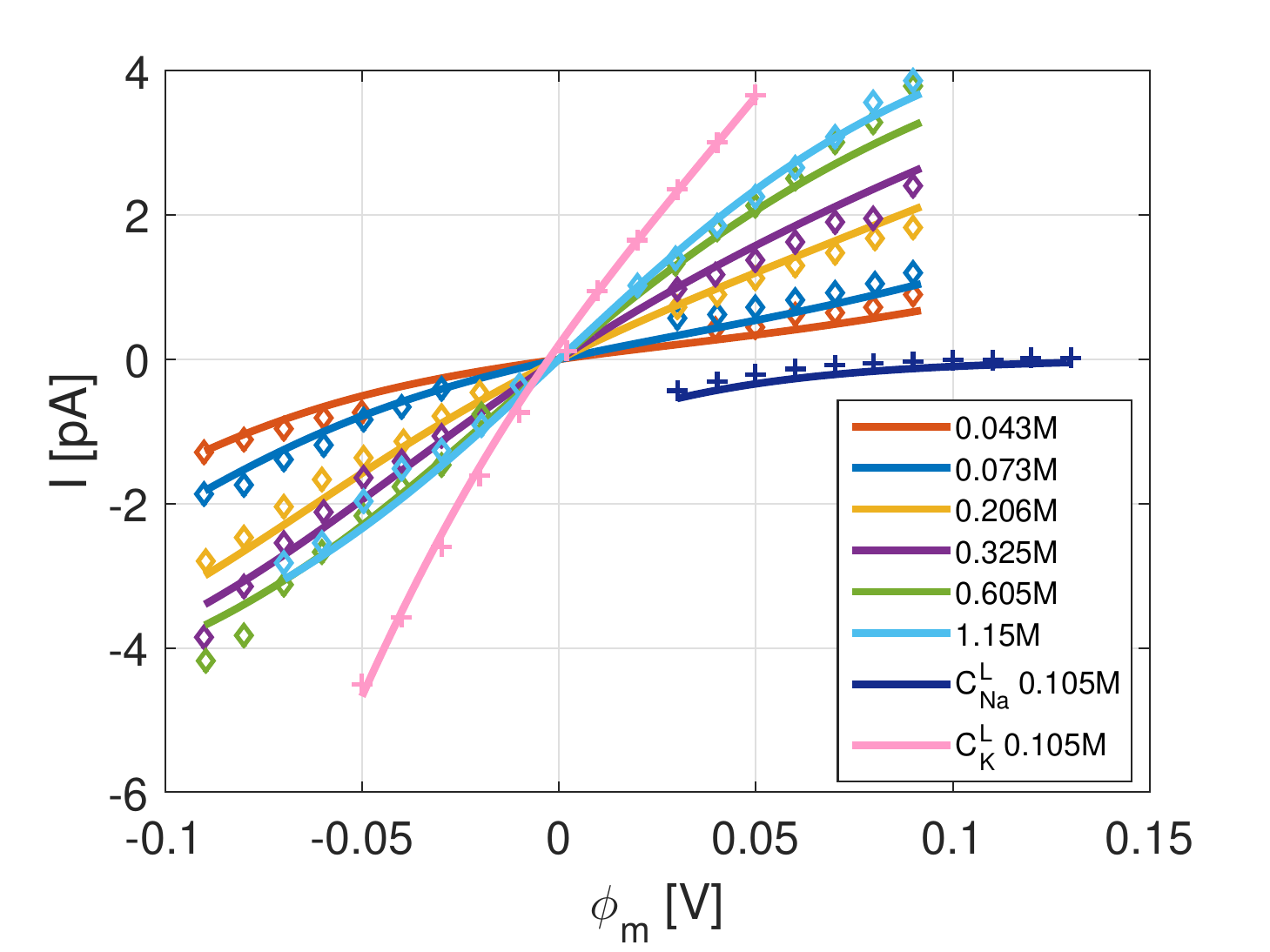}
\caption{Theoretical $I-V$ curves are compared with the experimental values for the Shaker K$^+$ channel \cite{Heginbotham:93}. Diamonds indicating symmetrical solutions, are fitted with the paramaters: $\tau_K=0.0486$, $\eta=0.617$ and $\Delta \tilde{\mu}^b_K=3.92kT$. The assymetrical solutions denoted by crosses are fitted with: $\tau_i=0.102 $, $\eta=0.629$,  and $\Delta \tilde{\mu}^b_i=4.14kT$.}\label{Fig:FittingShaker}
\end{center}
\end{figure}

In  Fig: \ref{Fig:FittingShaker} we display the results of comparisons with the Shaker channel, noting that the diamonds indicate symmetrical solutions and crosses the asymmetrical solutions, with concentrations given in the legend.  The asymmetrical solutions  consisted of   K$^+$ in the extracellular solution  with concentration  $0.095$M, and  either 0.105M K$^+$ or Na$^+$  in the intracellular solution. 

\begin{figure}[h!]
\begin{center}
\includegraphics[scale=0.585]{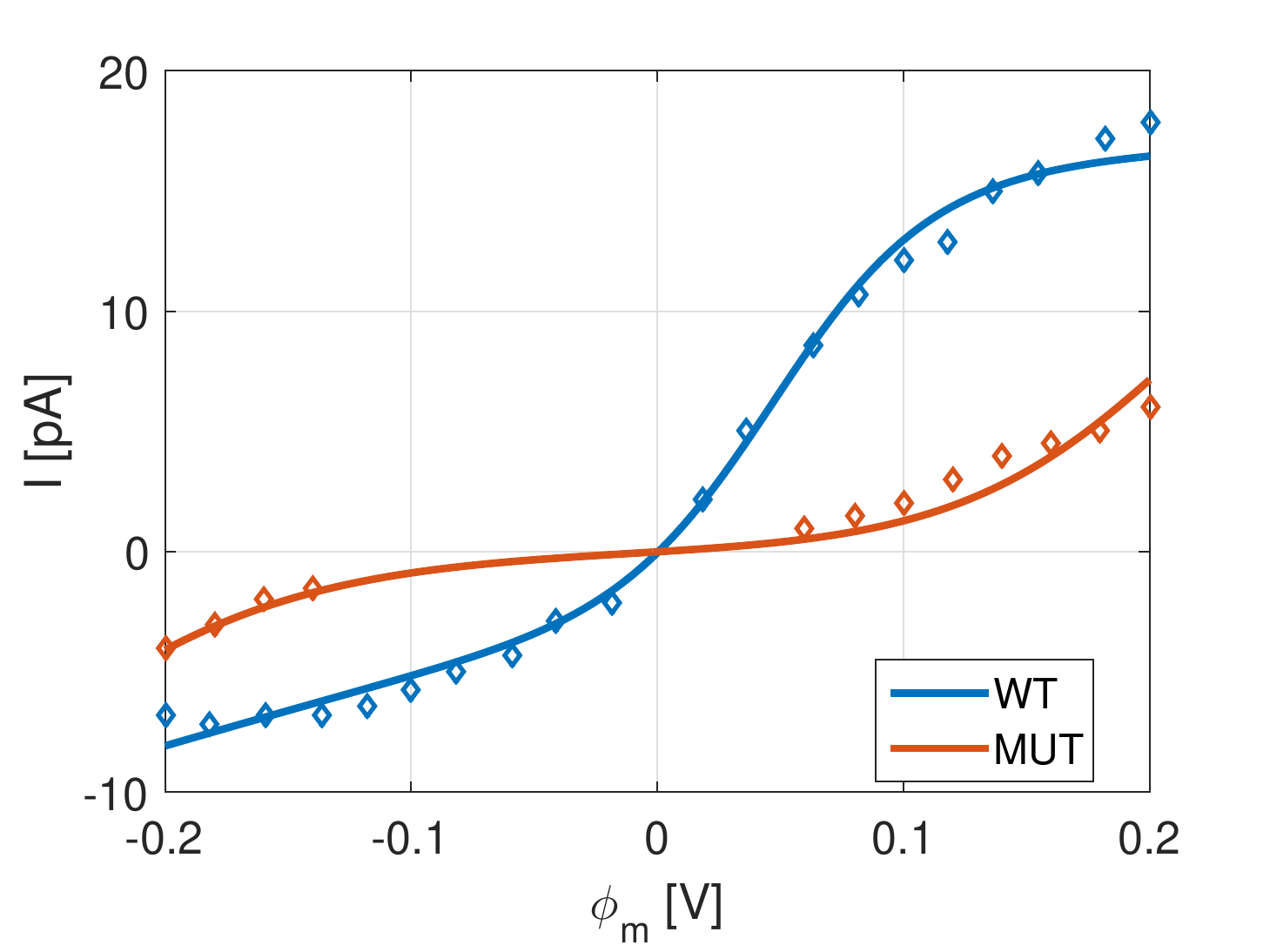}
\caption{Theoretical $I-V$ curves are compared with the experimental values for WT (blue) and MUT (orange) \cite{zhou:2004}. Resulting in: $\tau_K=0.161$, $\Delta \tilde{\mu}^b_{WT}=4.1kT$, $\Delta \tilde{\mu}^b_{MUT}=0.979kT$, $\eta_{WT}=0.175 $ and $\eta_{MUT}=0.446$.}\label{Fig:FittingMutant}
\end{center}
\end{figure}

The fitting parameters were very similar, although we note that the diffusion rate doubled when we had asymmetrical solutions. Importantly, however it was much less than the bulk value in agreement with simulations \cite{Tieleman:01}.  The remaining fitting parameters are consistent with  both experiments for, $\eta \sim 0.6$ and $\Delta \tilde{\mu}^b_K\sim 4kT$. This indicates a slight asymmetry in the channel and a low energy barrier for entry. The theory fits better at low concentrations possibly  due to the break-down of the Debye-H{\"u}ckel interaction term, which works best for dilute solutions.

In the final comparison we compare our theory to wild-type (WT) KcsA and its mutant (MUT), created by  replacing the  amino-acid threonine with cysteine at the S4 site (for full mutagenisis details refer to \cite{zhou:2004}), strongly affecting  conduction. To model this mutation  we allow for differences in symmetry and binding, and hence vary $\Delta \tilde{\mu}^b_K$ and $\eta$, between the WT and MUT. In Fig: \ref{Fig:FittingMutant} we display the results, and there is a strong difference in the symmetry and $\Delta \tilde{\mu}^b_K$ . In WT KcsA the energy barrier for entry was much smaller because $\Delta \tilde{\mu}_K^b$ was much larger than the mutant, and tuned to produce faster conduction. The difference in the $\eta$ parameters were large suggesting the site mutation caused an asymmetry from the original WT.

\section{Conclusion}
We have introduced a self-consistent kinetic model describing transitions of state in the selectivity filter of narrow channels. It demonstrates large selectivity amongst mono-valent species coexisting with high conductivity, and  CB effects. Predictions of the model are in good agreement with experimental data and we have considered the effect of a filter mutation. It can be extended by considering site-specific binding energies to increase the number of states, and with  further comparison to experimental data. 

We fully expect the theory to be applicable to other narrow channels such as voltage-gated sodium channels e.g. NaChBac, and to artificial nano-pores. 

\section*{Acknowledgements}
We are grateful to Bob Eisenberg, Miroslav Barabash, Aleksandra Pidde and Aneta Stefanovska for helpful discussions. The research was supported by the Engineering and Physical Sciences Research Council UK (grant No. EP/M015831/1).

\bibliographystyle{IEEEtran}
\bibliography{ionchannels--}

\end{document}